\documentclass[notitlepage]{revtex4-1}
\usepackage{amsmath}
\usepackage{amssymb}
\usepackage{natbib}
\usepackage{subfigure}
\usepackage{mathtools}
\usepackage{graphicx}
\usepackage{color}
\usepackage{setspace}
\usepackage{mathrsfs}
\usepackage[usenames,dvipsnames]{xcolor}
\def\JP#1{\textcolor{Black}{#1}}
\def\RK#1{\textcolor{Black}{#1}}

\newcommand\dd{\mathrm{d}}

\newcommand{\Z}{\mathbb Z}

\def\bdot{\raise.2em\hbox to .15em{.}}

\definecolor{gray}{gray}{0.5}

\def\bdotblack{\raise.25em\hbox to .15em{.}}

\begin{document}
\title{\RK{Koopman analysis of Burgers equation}}

\author{Jacob Page$^{*,\dag}$} 
\author{Rich R. Kerswell$^{*,\dag}$}
\affiliation{$^*$School of Mathematics, University of Bristol, Bristol, BS8 1TW, UK.}
\affiliation{$^\dag$Centre for Mathematical Sciences, University of Cambridge, Cambridge, CB3 0WA, UK.}

\date{\today}

\begin{abstract}

The emergence of Dynamic Mode Decomposition (DMD) as a practical way to attempt a Koopman mode decomposition of a nonlinear PDE presents exciting prospects for identifying invariant sets and slowly decaying transient structures buried in the PDE dynamics. However, there are many subtleties in connecting DMD to Koopman analysis and it remains unclear how realistic Koopman analysis is for complex systems such as the Navier-Stokes equations.  With this as motivation, we present here a \RK{full Koopman decomposition for the velocity field in Burgers equation by deriving explicit expressions for the Koopman modes {\em and} eigenfunctions - the first time this has been done for a nonlinear PDE.  The decomposition highlights the fact that different observables can require different subsets of Koopman eigenfunctions to express them and presents a nice example where: (i) the Koopman modes are linearly dependent
and so cannot be fit {\em a posteriori} to snapshots of the flow without knowledge of the Koopman eigenfunctions; and (ii) the Koopman eigenvalues are highly degenerate which means that computed Koopman modes become initial-condition dependent. As way of illustration, we discuss the form of the Koopman expansion  with various initial conditions and assess the capability of DMD to extract the decaying nonlinear coherent structures in run-down simulations. }

\end{abstract}

\maketitle


In complex systems such as turbulent fluid flows, it has long been hoped that the state of the system could be decomposed into a series of (spatially) simpler states with known time behaviour to aid understanding and offer opportunities for prediction and control. While this seems to go against the nonlinearity inherent in complex systems,   
the Koopman operator \citep{Koopman1931, Mezic2005}, which is an infinite dimensional {\em linear} operator that shifts {\em observables} (functionals) of the state forward in time, offers some reasons for optimism. Because of its linearity, the Koopman operator possesses eigenfunctions with exponential time behaviour (given by the associated eigenvalue)  which appear to capture some essence of the underlying PDE. In particular, neutral eigenfunctions can identify basins of attraction of invariant sets in the underlying dynamical system, while slowly decaying Koopman eigenfunctions identify least damped coherent structures in transient systems. 

However, until recently, it has been unclear how to analyse the Koopman operator except in the simplest settings. The discovery of Dynamic Mode Decomposition (DMD) \citep{Schmid2010} has changed this situation by presenting a practical way to extract \emph{dynamic modes} which have a fixed spatial structure and an exponential dependence on time from numerical and experimental time series \citep[e.g.][]{Bagheri2013,Jovanovic2014,DMDkutz}. These modes are (right) eigenvectors of a best-fit linear operator which maps between equispaced-in-time vectors of observables (measurements) from the dataset and can be connected under certain circumstances to (vector) Koopman {\em modes} which weight the (scalar) Koopman eigenfunctions in their expansion of the observable vector \citep{Rowley2009,Tu2014}. Beyond the considerable interest in its connection to the Koopman operator, DMD has rapidly become an invaluable tool in its own right for post-processing numerical and experimental time series \cite{DMDkutz,Rowley2017}.

\JP{
    The input to the DMD algorithm is a matrix whose $i^{th}$ column is a vector ${\boldsymbol \psi}_i$ of measurements (user-specified observables) of the dynamical system taken at time $t=t_i$.
    Typically the raw state vector $\boldsymbol \psi = \mathbf u$ is chosen (e.g. the velocity field in fluid flow problems).
    In this instance, connecting the output of DMD to the Koopman operator is fraught with uncertainty, since Koopman eigenfunctions are typically nonlinear functions of the state \citep[see e.g.][]{Rowley2017}.
    In response, several alternative data processing techniques have been proposed to extract Koopman modes from simulation/experimental data. 
    For example, \citet{Sharma2016} demonstrated that a purely oscillatory Koopman mode with frequency $m$ (a Fourier component of a limit cycle) is well described by the first response mode of the resolvent operator at the same frequency, 
    while \citet{Arbabi2017} have adapted techniques from signal processing to extract Koopman modes from chaotic, statistically stationary flows.
    \citet{Williams2015} modified the DMD algorithm (``extended dynamic mode decomposition'' -- EDMD) to make it better suited to identifying Koopman modes by populating the vectors $\boldsymbol \psi_i$ with \emph{functionals} of the state $\mathbf u$.  
    EDMD is a robust method for determining unknown Koopman eigenfunctions provided that (i) they can be expressed as a linear combination of the user-specified observables that make up the $\boldsymbol \psi_i$ and (ii) sufficient data is available \citep{Tu2014,Williams2015,Rowley2017}.
}

\JP{
    Choosing a suitable set of functionals for $\boldsymbol \psi$ is non-trivial, and is made more challenging by the fact that
    Koopman eigenfunctions have only been determined analytically for low-dimensional ODEs \RK{\citep[e.g.][] {Bagheri2013,Brunton2016,Rowley2017}}.
    Perhaps the most successful example is \citet{Bagheri2013}, who derived a Koopman decomposition for the Stuart-Landau equation to describe the transient collapse of flow past a cylinder onto a limit cycle.
    However, often in practice DMD is applied to very large dynamical systems ($O(10^6)$ degrees of freedom) with an underlying PDE where little is known about the Koopman eigenfunctions. 
    Moreover, even if a finite subset of the Koopman eigenfunctions can be determined, there is no guarantee (and it is often unlikely) that the state variable itself will lie in the span of this invariant subspace \citep{Brunton2016}. 
}

\RK{Given this backdrop, we provide the first complete Koopman analysis of a nonlinear PDE in order to examine the DMD-Koopman operator connection in a more realistic setting than has been done so far in ODEs.}  By exploiting the Cole-Hopf transformation \citep{Hopf1950,Cole1951}, we derive the full Koopman mode  decomposition of solutions to Burgers equation, including explicit analytical expressions for the Koopman eigenfunctions, and assess the capability of DMD to extract these objects from time series.
The Koopman mode expansion found here has a number of intriguing features which could have  implications for the interpretation of the outputs of DMD \RK{in more complex systems.}

%
%
\begin{figure}
    \centering
    \includegraphics[width=0.3\textwidth]{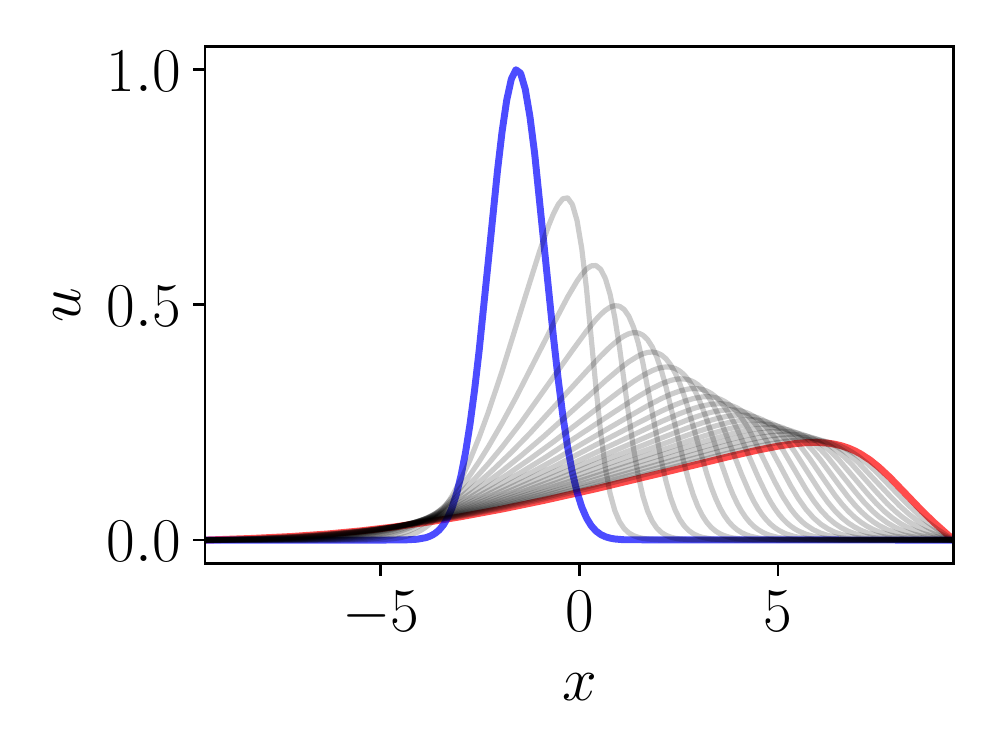}
    \includegraphics[width=0.3\textwidth]{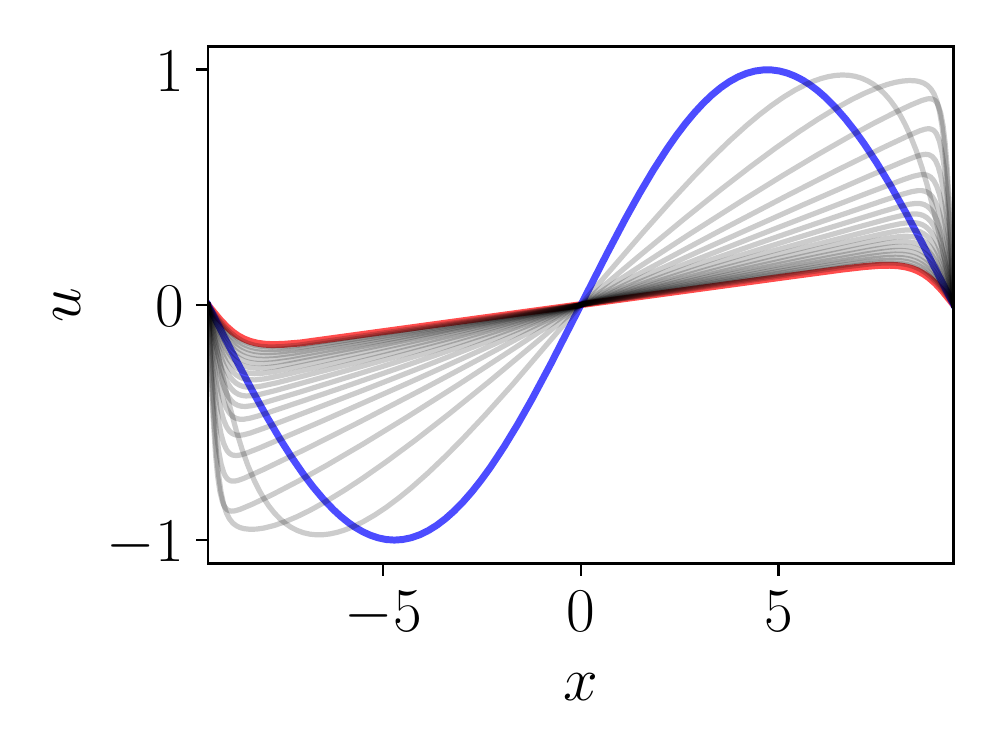}
    \vspace{-4mm}
    \caption{Evolution of (left) $u_G(x) = \text{exp}[-(x+\pi/2)^2]$ and (right) $u_S(x) = \text{sin}(2\pi x/L)$ for $t\leq 40$ (blue/red profile is the initial/final state).}
    \label{fig:u_evoln}
\end{figure}

We consider square-integrable solutions $u(x,t)$ to the one-dimensional Burgers equation,
\begin{equation}
    \partial_tu = F(u):= -u\partial_x u +\nu\partial^2_xu,
    \label{eqn:burgers}
\end{equation}
over a domain of length $L$ with Dirichlet conditions $u(\pm L/2,t)=0$.
Throughout, $L=6\pi$ and $\nu=0.1$ which is equivalent to a Reynolds number $Re=60\pi$ on a unit-length domain. Three initial conditions will be discussed in the text: a localised Gaussian pulse, $u_G(x) = \text{exp}[-(x + \pi/2)^2]$, a shifted pulse $u_G(x-\pi/2)$ and a sine wave, $u_S(x) = \text{sin}(2\pi x/L)$: see figure \ref{fig:u_evoln}. 

If the time-forward map of (\ref{eqn:burgers}) is $f^t(u) := u + \int_0^tF(u)\dd t'$, then the Koopman operator is defined by
\begin{equation}
\mathscr K^t \boldsymbol{\psi }(u) := \boldsymbol{\psi}(f^t(u))
\end{equation}
where $\boldsymbol{\psi} \in \mathbb C^N$ is a vector made up of $N$ observables of the state vector $u$ \RK{which simply could just be one functional of $u(x,t) $ evaluated at different $x$} . A Koopman eigenfunction $\varphi$ is a special observable for which $\mathscr K^t \varphi(u) = \varphi(f^t(u)) =\varphi(u) e^{\lambda t}$ ($\lambda$ being the eigenvalue) and \RK{(assuming completeness)} the set of all Koopman eigenfunctions $\{\varphi_n\}$ can be used to represent and time-advance a vector of observables, $\boldsymbol{\psi}(u)$ as follows
\begin{equation}
\mathscr K^t  \boldsymbol{\psi} (u) =
  \mathscr K^t \sum_{n=0}^{\infty} \varphi_n(u) \hat{\boldsymbol \psi}_n
  =
 \sum_{n=0}^{\infty} \varphi_n(u)e^{\lambda_n t} \hat{\boldsymbol \psi}_n
\end{equation}
where $\{ \hat{\boldsymbol{\psi}}_n\}$ are the (vector) Koopman modes specific to the vector of observables under consideration. The apparent linearity in this expansion is a consequence of the coordinates; in general an individual mode, \RK{$e^{\lambda_n t}\hat{\boldsymbol \psi}_n$ cannot be considered in isolation. For example, if ${\boldsymbol \psi}$ is defined as simply $u$ evaluated at different $x$ values, $e^{\lambda t} \hat{\boldsymbol \psi}_n$ is {\em not} a (discretized) solution of the governing equation (\ref{eqn:burgers}).}

%
%
\JP{For nonlinear systems the requirements for DMD and Koopman analysis to overlap are challenging in practice to enforce. In linear systems, however, the Koopman eigenfunctions are linear functionals of the state and DMD with the full state variable will yield all of the Koopman modes \citep{Rowley2017}.} Fortunately, the well-known Cole-Hopf transformation \citep{Hopf1950,Cole1951}
\begin{equation} 
u: =-2\nu \partial_x v/ v, \quad
v:= e^{\xi(u)}\biggl/\int^{L/2}_{-L/2} e^{\xi(u)}  \, \dd x \biggr.
\label{CH}
\end{equation}
where
$\xi(u):=-\frac{1}{2 \nu} \int^x_{-L/2} u(x',t) \, \dd x'$
converts Burgers equation for $u$ with $u(\pm L/2,t)=0$ into the simple diffusion equation $\partial_t v=\nu \partial^2_x v$ for $v(x,t)$ with $\partial_x v(\pm L/2,t)=0$. 
\JP{The utility of the Cole-Hopf transform in the context of a Koopman analysis was noted in \citet{Kutz2016}.
As described below, it can be leveraged to obtain Koopman eigenfunctions due to the fact that $v(u;x)$ can be treated as an array of observables (functionals of $u$) parameterized by $x$ whose evolution is governed by a linear equation. 
Since the diffusion equation is highly tractable, an analytical spectral analysis of the associated Koopman operator is possible.
}

%
%
A Koopman mode expansion of $v(u;x)$ takes the form
\begin{equation}
    v(u;x) = \sum_{n=0}^{\infty} \varphi_n(u)\hat{v}_n(x)
    \label{Koopman_v}
\end{equation}
where the dependence on $u$ (via the Koopman eigenfunctions $\varphi_n$) and $x$ (via the Koopman modes $\hat{v}_n$) is separated. Given $\mathscr K^t \varphi_n(u) =\varphi_n(u) e^{\lambda_n t}$ by definition and the linearity of the diffusion equation, the Koopman modes $\hat{v}_n(x)$ have to be eigenfunctions of the 1D Laplacian. This together with the boundary conditions means that the only possibility is that  $\hat{v}_n = \text{cos}(n\pi x/L)$ ($n$ even) or $\hat{v}_n = \text{sin}(n\pi x/L)$ ($n$ odd). The Koopman eigenfunctions required to evolve $v$ are then
\begin{equation}
    \varphi_n(u) = \langle \hat{v}_n(x), v(u; x) \rangle, 
    \label{eqn:koopman_v}
\end{equation}
where $\langle f, g\rangle := \frac{2}{L} \int^{L/2}_{-L/2}\, f g\, \dd x$, so that the observable $v$ can be evolved in time as follows: $\mathscr K^t v(u;x) = v(f^t(u); x) = \sum_n \varphi_n(u)e^{\lambda_n t} \hat{v}_n(x)$,
with the Koopman eigenvalues $\lambda_n = -n^2\pi^2 \nu /L^2$.

%
%
Now the  strategy is to leverage knowledge of the expansion (\ref{Koopman_v}) by exploiting the Cole-Hopf transformation to perform a Koopman analysis of $u$. We write
\begin{equation}
    u(x) = \sum_{n=0}^{\infty} \Phi_n(u) \hat{u}_n(x),
    \label{eqn:koopman_u}
\end{equation}
using different symbols for the Koopman eigenfunctions ($\Phi_l$) and eigenvalues ($\Lambda_l$) here since an expanded set (which includes the $\varphi_n(u)$ and $\lambda_n$ defined in equation (\ref{eqn:koopman_v})\,) will be required. At time $t$, the Cole-Hopf transformation in the form $uv=-2 \nu \partial_x v$ reads
\begin{align}
    \sum_{l=0}^{\infty}\sum_{m=0}^{\infty} 
    \Phi_l(u) \hat{u}_l(x) \, \varphi_m(u)\hat{v}_m(x) e^{(\Lambda_l-m^2\pi^2\nu/L^2) t} = 
            -2 \nu\sum_{n=0}^{\infty} \varphi_n(u) \frac{\dd \hat{v}_n(x)}{\dd x} e^{-n^2\pi^2\nu t/L^2}
\end{align}
which allows the Koopman eigenfunction and Koopman mode pairs to be determined from the recurrence relation
\begin{equation}
    \sum_{l+m^2=k} \Phi_l(u)\hat{u}_l \,\varphi_m(u)\hat{v}_m(x) = \begin{cases}
        -2\nu \varphi_n(u)\frac{\dd\hat{v}_n}{\dd x} & k=n^2 \\
        0 & k \neq n^2
    \end{cases}
    \label{eqn:constraint}
\end{equation}
where $m,n,l, k \in \{0\}\cup \mathbb N$, and by inspection, the Koopman eigenvalues are $\Lambda_l=-l \pi^2 \nu/L^2$.

\begin{figure}
    \centering
    \includegraphics[width=0.5\textwidth]{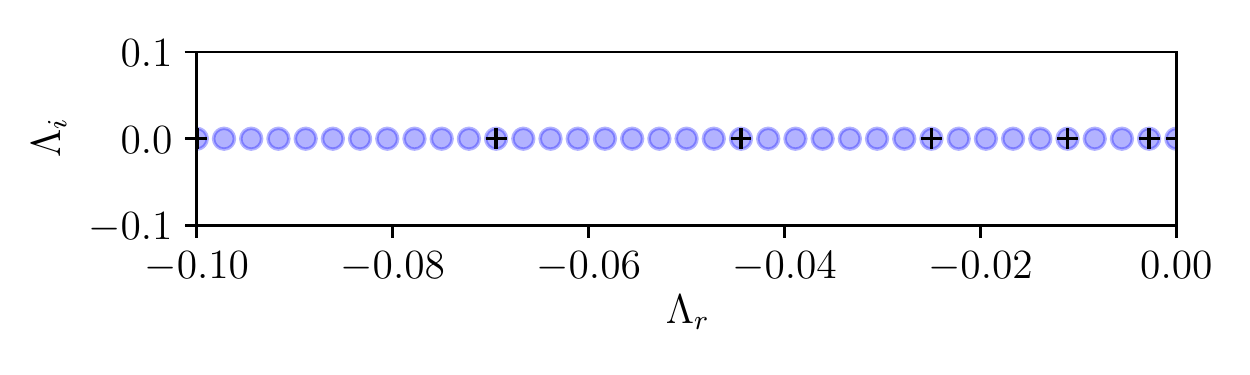}
    \includegraphics[width=0.5\textwidth]{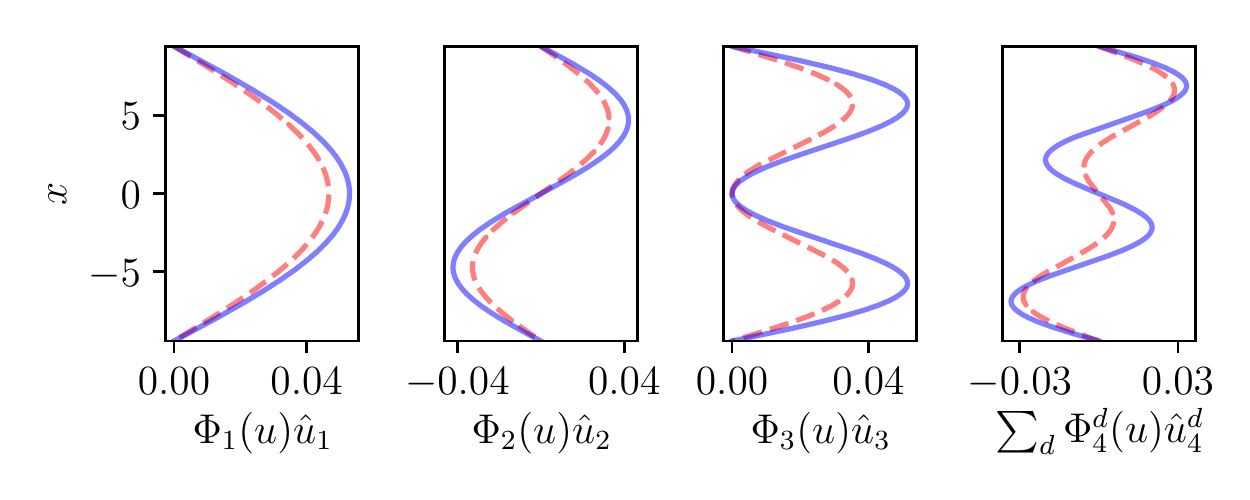}
    \includegraphics[width=0.5\textwidth]{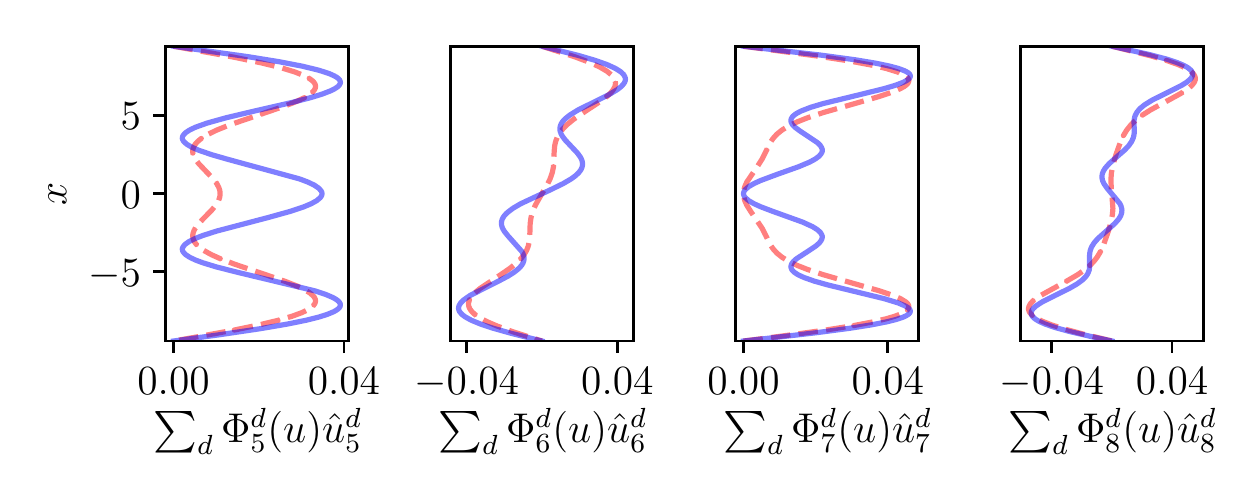}
    \vspace{-4mm}
    \caption{(Top) Koopman eigenvalues for Burgers equation (filled circles) with the eigenvalues required to evolve $v(u)$ overlaid as crosses. (Bottom) The first eight non-trivial Koopman modes, summed over their respective degeneracies, evaluated with $u=u_G$ (blue) and $u=u_G(x-\pi/2)$ (dashed red): \RK{The {\em shape} of Koopman modes 4 and above are dependent on initial conditions whereas that of modes 1-3 are not}.}    \label{fig:koop_eig}
\end{figure}

The Koopman eigenvalues and the first eight non-zero modes (evaluated with initial condition $u_G(x)$) are reported in figure \ref{fig:koop_eig} (the neutral mode $\Lambda_0=0$ is just $\hat{u}_0(x) = 0$ which is the only smooth equilibrium possible in the current configuration \citep{Benton1972}). 
\JP{
Analytical expressions for the first eight non-trivial pairs are
\begin{subequations}
    \begin{equation}
        \Phi_1(u)\hat{u}_1(x) = -\frac{2\pi\nu}{L}
        \left(\frac{\varphi_1(u)}{\varphi_0(u)}\right)\text{cos}(\pi x/L) \\
    \end{equation}
    \begin{equation}
        \Phi_2(u)\hat{u}_2(x) = \frac{\pi\nu}{L}
        \left(\frac{\varphi_1(u)}{\varphi_0(u)}\right)^2\text{sin}(2\pi x/L) \\
    \end{equation}
    \begin{equation}
        \Phi_3(u)\hat{u}_3(x) = -\frac{2\pi\nu}{L}
        \left(\frac{\varphi_1(u)}{\varphi_0(u)}\right)^3\text{sin}^2(\pi x/L)\text{cos}(\pi x/L) \\
    \end{equation}
    \begin{equation}
        \sum_{d=1}^2\Phi_4^d(u)\hat{u}_4^d(x) = \frac{\pi\nu}{L}
        \left(\frac{\varphi_1(u)}{\varphi_0(u)}\right)^4\text{sin}^2(\pi x/L)\text{sin}(2\pi x/L)
         + \frac{4\pi\nu}{L}\left(\frac{\varphi_2(u)}{\varphi_0(u)}\right)\text{sin}(2\pi x/L),
    \end{equation}
    \begin{align}
        \sum_{d=1}^2\Phi_5^d(u)\hat{u}_5^d(x) = &-\frac{\pi\nu}{L}
        \left(\frac{\varphi_1(u)}{\varphi_0(u)}\right)^5\text{sin}^3(\pi x/L)\text{sin}(2\pi x/L) \nonumber \\
        & + \frac{2\pi\nu}{L}\left(\frac{\varphi_2(u)\varphi_1(u)}{\varphi_0^2(u)}\right)\bigg[\text{cos}(\pi x/L)\text{cos}(2\pi x/L) 
    - 2 \text{sin}(\pi x/L)\text{sin}(2\pi x/L)\bigg],
    \end{align}
    \begin{align}
        \sum_{d=1}^2\Phi_6^d(u)\hat{u}_6^d(x) = &\frac{\pi\nu}{L}
        \left(\frac{\varphi_1(u)}{\varphi_0(u)}\right)^6\text{sin}^4(\pi x/L)\text{sin}(2\pi x/L) \nonumber \\
        & + \frac{\pi\nu}{L}\left(\frac{\varphi_2(u)\varphi_1^2(u)}{\varphi_0^3(u)}\right)\bigg[4 \text{sin}^2(\pi x/L)\text{sin}(2\pi x/L) 
    - \text{sin}(4\pi x/L)\bigg],
    \end{align}
    \begin{align}
        \sum_{d=1}^2\Phi_7^d(u)\hat{u}_7^d(x) = &-\frac{\pi\nu}{L}
        \left(\frac{\varphi_1(u)}{\varphi_0(u)}\right)^7\text{sin}^5(\pi x/L)\text{sin}(2\pi x/L) \nonumber \\
        & + \frac{\pi\nu}{L}\left(\frac{\varphi_2(u)\varphi_1^3(u)}{\varphi_0^4(u)}\right)\bigg[\frac{3}{2} \text{sin}(\pi x/L)\text{sin}(4\pi x/L)
    - 4\text{sin}^3(\pi x/L)\text{sin}(2\pi x/L)\bigg],
    \end{align}
    \begin{align}
        \sum_{d=1}^3\Phi_8^d(u)\hat{u}_8^d(x) = &\frac{\pi\nu}{L}
        \left(\frac{\varphi_1(u)}{\varphi_0(u)}\right)^8\text{sin}^6(\pi x/L)\text{sin}(2\pi x/L) -\frac{2\pi\nu}{L} \left(\frac{\varphi_2^2(u)}{\varphi_0^2(u)}\right)\text{sin}(4\pi x/L)\nonumber \\
        & + \frac{\pi\nu}{L}\left(\frac{\varphi_2(u)\varphi_1^4(u)}{\varphi_0^5(u)}\right)\bigg[4\text{sin}^4(\pi x/L)\text{sin}(2\pi x/L)
    - 2\text{sin}^2(\pi x/L)\text{sin}(4\pi x/L)\bigg].
    \end{align}
    \label{eqn:4modes}
\end{subequations}
}

%
%
%

\JP{
    Equations (\ref{eqn:4modes}\emph{d}-\emph{g}) indicate that the Koopman eigenvalues of Burgers equation are degenerate, and that the degree of degeneracy increases with the mode number.
    Of the modes reported above, numbers $4$ through $7$ are 2-fold degenerate while mode $8$ is 3-fold degenerate.
    Continuing upwards in mode number, it can be shown that mode $9$ is 4-fold degenerate ($9 = 9\times 1^2 = 2^2 + 5\times 1^2 = 2 \times 2^2+1^2 = 3^2$, and the corresponding eigenfunctions $\{\Phi_9^d\}$ are $\{\varphi_1^9,  \varphi_2\varphi_1^5, \varphi_2^2\varphi_1,\varphi_3\}$). 
    The cause of the degeneracy is connected to the fact that if $\varphi_i$ and $\varphi_j$ are Koopman eigenfunctions with respective eigenvalues $\lambda_i$ and $\lambda_j$, then $\varphi_i\varphi_j$ is also a Koopman eigenfunction with eigenvalue $\lambda_i+\lambda_j$ \citep{Mezic2013}. 
    Since the linear dynamics of $v(u)$ are governed by a simple diffusion, the associated eigenvalues $\{\lambda_n\}$ are those of the Laplacian and are proportional to perfect squares. 
    Therefore, the eigenvalue $\Lambda_l$ will be $p$-fold degenerate, where $p$ is the number of ways the perfect squares $\leq l$ can be combined to give $l$.
    In fact, it has been proved recently that  $p(l)\sim \exp(a\,l^{1/3})$ asymptotically (where $a$ is a positive constant) \citep{Vaughan15} and so grows very quickly with $l$.}

\RK{Despite the ubiquity of the Laplacian, this eigenvalue degeneracy is not generic since a more general  evolution operator based on a Laplacian will also contain other terms which will disrupt the ordered distribution of the eigenvalues.  For example, in the Navier-Stokes equations, the presence also of advection terms is enough to prevent this degeneracy. However, there is also reason to believe this degeneracy occurs in other nonlinear PDEs and so is not unique to Burgers either. For example, the KdV equation can be solved via an inverse scattering transform \citep{DrazinSoliton}. As part of this procedure, the initial condition on $u$ is treated as a potential from which scattering data are derived.    The scattering data (a set of observables) can be evolved linearly in time before the inverse scattering transform is used to infer the state variable at later times.    This procedure yields a subset of the Koopman eigenvalues.    For the classic $N$-soliton solution, these eigenvalues are proportional to the perfect cubes, allowing for the emergence of degeneracy in a manner analogous to Burgers equation. }

\JP{
The degeneracy in Burgers equation has important consequences for interpreting the results of DMD since a data set derived from a single initial condition can only carry information  about {\em one} realization from the multi-dimensional eigenspace associated with the eigenvalue. This means the dynamic mode which emerges from applying DMD will be initial-condition dependent contrary to usual expectations. This point is illustrated in figure \ref{fig:koop_eig}, where the Koopman modes (summed over their degeneracies) associated with the shifted initial condition $u_G(x-\pi/2)$ are overlaid onto those obtained with $u_G(x)$. For the non-degenerate modes ($n\in \{1,2,3\}$) the result is a trivial rescaling whereas for the degenerate modes $n\geq 4$, the shape functions are altered significantly. To circumvent this issue, DMD would have to treat data from at least $p$ different initial conditions to have any chance of properly sampling the $p$-dimensional eigenspace of a degenerate  Koopman eigenvalue.
}

The Koopman expansion for $u$ is a decomposition of the solution of a nonlinear PDE into a superposition of fixed nontrivial coherent structures with simple (exponential) temporal behaviour, but it would be a mistake to conclude that  a Koopman mode has dynamical significance in isolation. They are not individually solutions to (\ref{eqn:burgers}) and generally it is not possible to set up an initial condition which only excites a single Koopman mode. For example, while the initial condition $u_S(x)$ can be expressed solely in terms of the second Koopman mode, $\hat{u}_2(x)$, their time evolutions do not match. This is because the associated eigenfunction $\Phi_2(u_S)=0$ 
\footnote{Initial conditions invariant under $\Z: u(x) \rightarrow -u(-x)$ require only a subset of $\{\Lambda_l\}$ to describe their evolution.}
so $u_S(x)$ has to be expressed using an infinite sum of all the other Koopman modes. Moreover, the Koopman modes are \emph{linearly dependent}: compare the Koopman mode associated with $\Lambda_2$ (\ref{eqn:4modes}\emph{b}) to the second Koopman mode associated with the 2-fold degenerate $\Lambda_4$ (\ref{eqn:4modes}\emph{d}).
This linear dependence has important consequences for interpreting the results of DMD which are often used to build a low-dimensional model of the dynamics \citep[e.g.][]{DMDkutz}. Normally a least-squares fit is performed on a subset of DMD modes to a snapshot of the system but this is clearly inappropriate for a linearly dependent set of Koopman modes.  Correct `fitting' can only be guaranteed by evaluating the Koopman eigenfunctions $\{\Phi_n(u)\}$, which uniquely determine the amplitudes of the Koopman modes.

%
%
\begin{figure}
    \centering
    \includegraphics[width=0.3\textwidth]{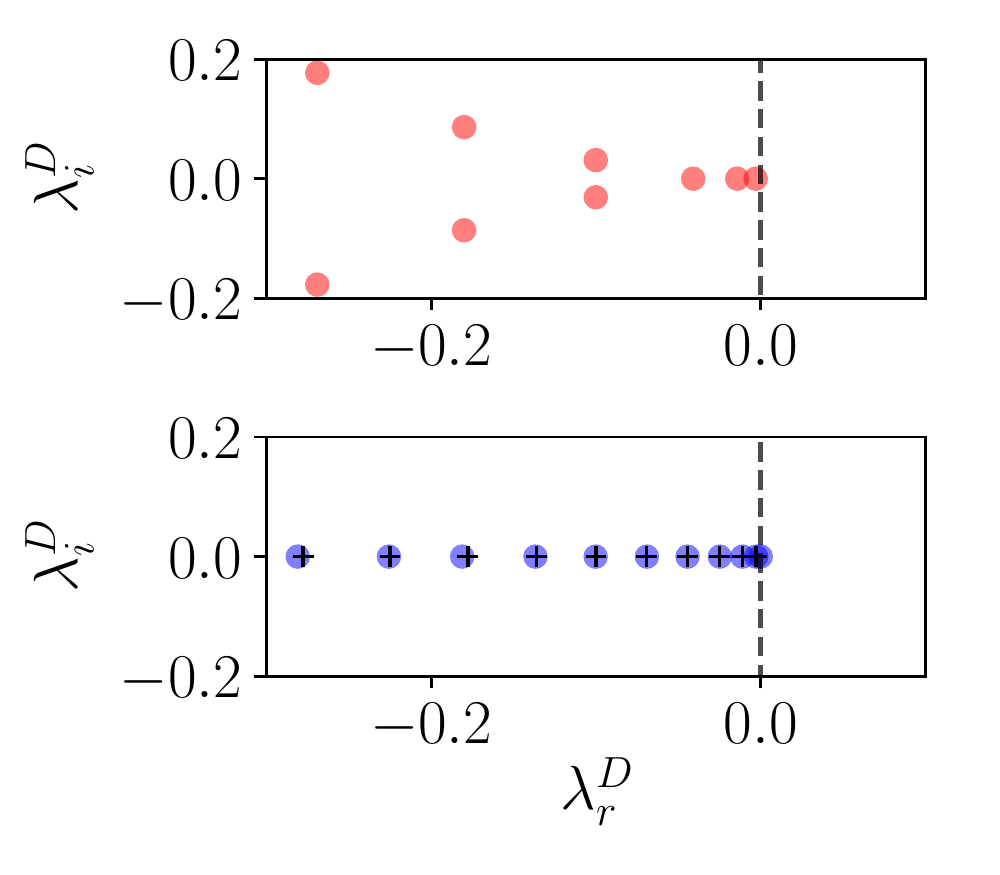}
    \includegraphics[width=0.3\textwidth]{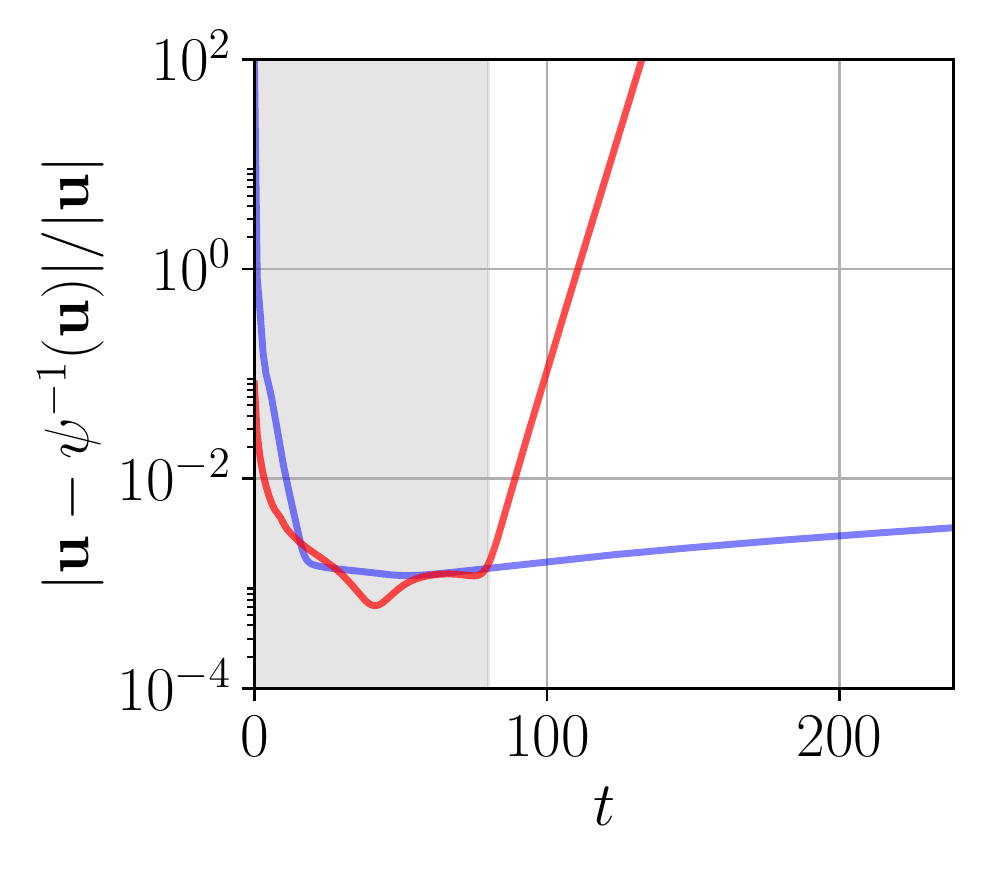}
    \includegraphics[width=0.3\textwidth]{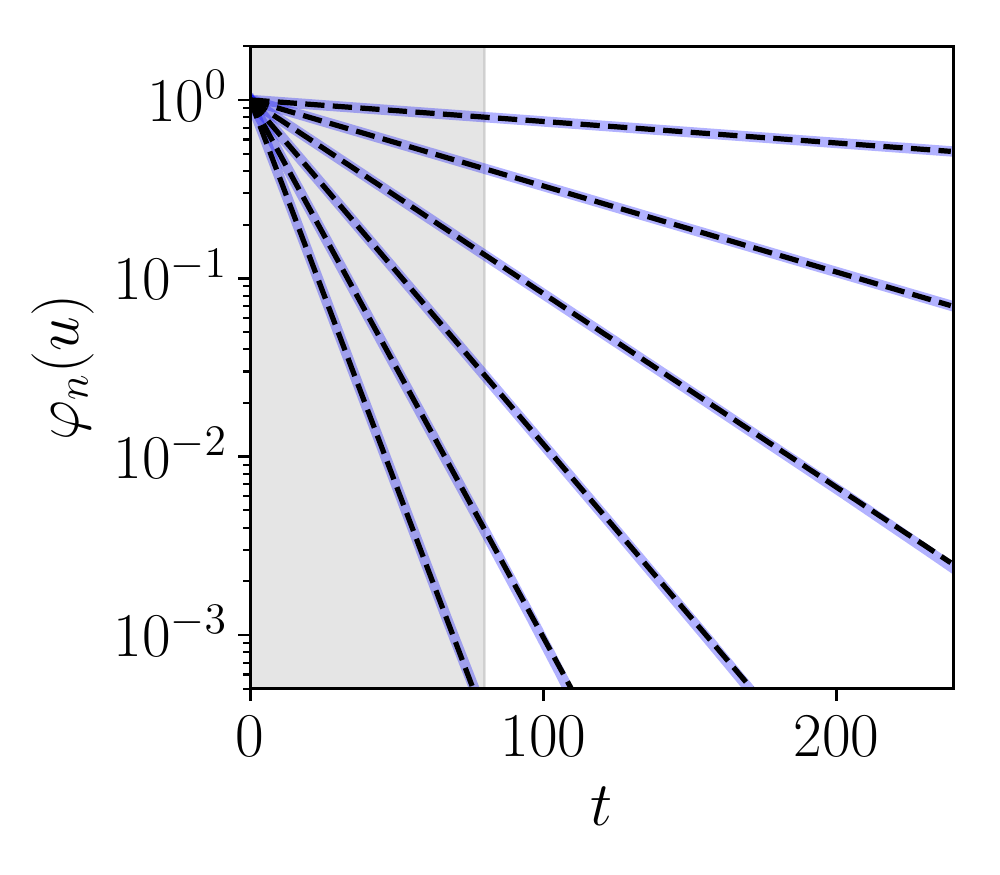}
    \vspace{-4mm}
    \caption{(Left) Eigenvalues from DMD of $u_G$ with $M=200$ snapshot pairs with spacing $\delta t=1$, collected within the time window $t\in [0,80)$ for both $\boldsymbol \psi = \mathbf u$ (red) and $\boldsymbol \psi = \mathbf v(\mathbf u)$ (blue). Note the crosses are the eigenvalues of the linear diffusion problem $\lambda_n = -n^2\pi^2 \nu/L^2$. The rank of the problem $r=70$ for $\boldsymbol \psi=\mathbf u$ and $r=45$ for $\boldsymbol \psi =\mathbf v(\mathbf u)$. (Centre) The error in the approximation $\boldsymbol \psi^{-1}\left(\sum_j a_j \hat{\boldsymbol \psi}^D_j\text{exp}(\lambda_j t)\right)$ as compared with the state vector $\mathbf u$, evaluated beyond the observation window (shown in grey; colors as in left panel). The $a_j$ are determined from a least-squares fit for the observable over the DMD window. 
        \JP{(Right) Koopman eigenfunctions $\{\varphi_n(u)\}$ ($n=1,\dots,6$) required to evolve $v(u)$: blue lines are direct evaluation of $\varphi_n(u) = \langle \hat{v}_n, v(u) \rangle$ on the trajectory $f^t(u_G)$; dashed, black lines are Koopman eigenfunctions obtained from the DMD with $\boldsymbol \psi = \mathbf v(\mathbf u)$}.}
    \label{fig:DMD}
\end{figure}


If the Koopman mode expansion in  (\ref{eqn:koopman_u}) is to be used to  evolve
$u$ forward in time, a large number of Koopman modes may need to be accurately resolved. In fact, for both $u_G$ and $u_S$, the Koopman eigenfunctions $\Phi_n$ decay only very slowly as $n\to \infty$ (sharp $v$-variations at $t=0$ translate into a large number of Koopman modes in $u$). The near-continuum of distinct timescales (the dense eigenvalues $\{\Lambda_l\}$ in figure \ref{fig:koop_eig}) also renders selection of the adjustable parameters in DMD difficult (time window length, snapshot spacing and potentially a rank truncation of the data matrix). In general, these selection issues require one to focus on a particular timescale of interest -- e.g. in run-down Burgers' calculations, the focus may be on the most slowly decaying Koopman modes
\footnote{At large times, $u_G \to \Phi_1(u_G)\text{cos}(\pi x/L)\text{exp}(-\pi^2\nu t/L^2)$ and $u_S \to \Phi_4(u_S)\text{sin}(2\pi x/L)\text{exp}(-4\pi^2\nu t/L^2)$}. 

As way of illustration, the output of a DMD of the propagating and decaying Gaussian pulse, $u_G(x)$, is reported in figure \ref{fig:DMD} (using the algorithm in \citep{Tu2014}). 
The data was obtained from numerical solution of Burgers' equation using an expansion in $N=250$ Chebyshev polynomials (hence the state vector $\mathbf u \in \mathbb R^N$, where $\mathbf u \cdot \mathbf e_i = u_G(x_i)$).
Two observables are considered: the raw state vector $\boldsymbol \psi(\mathbf u) = \mathbf u$ and the linearizing observable $\boldsymbol \psi(\mathbf u) = \mathbf v(\mathbf u)$.
Clearly, the eigenvalues in the former case do not match the analytical solution which are all on the negative real axis \JP{\citep[the displacement of the eigenvalues from the negative real axis shares qualitative similarities with DMD results reported in][]{Bagheri2013}.}
There is also a rapidly growing error when the DMD approximation is advanced beyond the observation window.
In this case, the failure to resolve the Koopman eigenfunctions means that DMD here amounts to an elaborate curve fitting exercise.

\JP{
    The calculation performed with the observable $\boldsymbol \psi(\mathbf u) = \mathbf v(\mathbf u)$ is an example of EDMD \citep{Williams2015}.
    Here, the elements of the observable vector constitute a suitable basis for the eigenfunctions $\{\varphi_n(u)\}$ and thus DMD successfully extracts the exact decay rates for the linearizing observable $\mathbf v$.
    However, note that the full state vector cannot be expressed solely in terms of these eigenfunctions, and so the EDMD algorithm would fail to extract Koopman modes for $\mathbf u$ without further augmentation of $\boldsymbol \psi$. 
Instead, the Cole-Hopf transform is applied to a superposition of EDMD modes for $\mathbf v$ to obtain the $\mathbf u$-evolution, and the correct identification of the slow Koopman modes for $\mathbf v$ is highlighted in figure \ref{fig:DMD} via the agreement between this approximation} and the original trajectory far beyond the observation window. 
\JP{
    In addition, figure \ref{fig:DMD} also includes a comparison of the DMD approximation to the first few Koopman eigenfunctions ($\mathbf w_j^H\mathbf v(\mathbf u)$, where $\mathbf w_j$ is the $j^{th}$ left-eigenvector of the DMD operator) to the analytical result.
    Again, excellent agreement is observed far beyond the observation window.
}
However, note that the practical issue highlighted above applies here:
while the parameters of the DMD were sufficient to recover the slow dynamics of the system, the failure to resolve the rapidly decaying Koopman modes is apparent in the significant error at early times \RK{ (middle plot of figure \ref{fig:DMD})}.

%

In this \RK{communication we have analytically derived the full Koopman mode and eigenfunction decomposition for Burgers equation -- the first time this has been done for a nonlinear PDE. This has provided a nice example of how  different sets of Koopman eigenfunctions are needed to propagate different observables (e.g. compare the set of Koopman eigenvalues for $\psi=u$ and $\psi=v$). It has also revealed   Burgers equation as an accessible nonlinear PDE in which  the Koopman eigenvalues are highly degenerate and Koopman modes can be linearly dependent. This system, therefore, should prove an invaluable testing ground for the refinement of DMD/Koopman schemes  designed to tackle these subtleties in more complex systems. }

\emph{Acknowledgements:} This work was funded by EPSRC under grant EP/K034529/1. The authors thank T. Wooley for bringing \citep{Vaughan15} to their attention \RK{and the referees for their helpful comments which improved the manuscript}.

\bibliography{koopman}
\end{document}